\documentclass[sigconf]{acmart}

\copyrightyear{2026}
\acmYear{2026}
\setcopyright{cc}
\setcctype{by}
\acmConference[MSR '26]{23rd International Conference on Mining Software Repositories}{April 13--14, 2026}{Rio de Janeiro, Brazil}
\acmBooktitle{23rd International Conference on Mining Software Repositories (MSR '26), April 13--14, 2026, Rio de Janeiro, Brazil}
\acmPrice{}
\acmDOI{10.1145/3793302.3793599}
\acmISBN{979-8-4007-2474-9/2026/04}

\usepackage{url}
\usepackage{multirow}
\usepackage{subcaption}
\usepackage{framed}
\colorlet{shadecolor}{gray!15}

\newenvironment{graybox}{%
  \MakeFramed{\advance\hsize-\width\FrameRestore}%
}{\endMakeFramed}

\title{On the Footprints of Reviewer Bots' Feedback on Agentic Pull Requests in OSS GitHub Repositories}

\author{Syeda Kaneez Fatima}
\email{s.fatima@lums.edu.pk}
\affiliation{%
  \institution{Lahore University of Management Sciences}
  \city{Lahore}
  \state{Punjab}
  \country{Pakistan}
}

\author{Yousuf Abrar}
\email{27100131@lums.edu.pk}
\affiliation{%
  \institution{Lahore University of Management Sciences}
  \city{Lahore}
  \state{Punjab}
  \country{Pakistan}
}

\author{Abdul Rehman Tahir}
\email{25030003@lums.edu.pk}
\affiliation{%
  \institution{Lahore University of Management Sciences}
  \city{Lahore}
  \state{Punjab}
  \country{Pakistan}
}

\author{Amelia Nawaz}
\email{27100025@lums.edu.pk}
\affiliation{%
  \institution{Lahore University of Management Sciences}
  \city{Lahore}
  \state{Punjab}
  \country{Pakistan}
}

\author{Shamsa Abid}
\email{shamsa.abid@lums.edu.pk}
\affiliation{%
  \institution{Lahore University of Management Sciences}
  \city{Lahore}
  \state{Punjab}
  \country{Pakistan}
}

\author{Abdul Ali Bangash}
\email{abdulali@lums.edu.pk}
\affiliation{%
  \institution{Lahore University of Management Sciences}
  \city{Lahore}
  \state{Punjab}
  \country{Pakistan}
}

\date{\today}

\begin{document}

\begin{abstract}
Autonomous coding agents are reshaping software development by creating pull requests (PRs) on GitHub, referred to as agentic PRs. In parallel, the review process is also becoming autonomous, thereby making reviewer bots key actors in the assessment of these agentic PRs. However, their influence on PR acceptance and resolution remains unclear. This study empirically investigates the relationship between reviewer-bot feedback and PR outcomes by analyzing how \textit{Reviewer Bot Feedback Quality} (relevance, clarity, conciseness) and \textit{Reviewer Bot Activity Volume} (comment count) are associated with PR acceptance and resolution time. We analyze $7,416$ reviewer-bot comments on $4,532$ PRs from the AI\_Dev dataset (a dataset that captured AI agents' PRs in GitHub projects). Our results show that reviewer-bot comments mainly focus on bug fixes, testing, and documentation, are civil in tone, and are prescriptive in nature. Reviewer bots generally produce clear and concise feedback, though the semantic relevance of comments to underlying code changes is moderate. We find that higher \textit{Reviewer Bot Activity} volume is associated with longer PR resolution times and lower average feedback quality, showing that as bots generate more comments on a PR, the average pertinence of that feedback appears to degrade. At the same time, \textit{Reviewer Bot Feedback Quality} shows no meaningful association with workflow outcomes. Our findings suggest that, in agentic PR workflows, reviewer bots should prioritize targeted high-relevance feedback over generating large numbers of comments.
\end{abstract}



\maketitle
\section{Introduction}
Autonomous coding agents introduce a distinct class of pull requests (PRs) in which code is generated or modified entirely by AI systems \cite{watanabe2025useagenticcodingempirical}. We refer to these contributions as \textit{agentic pull requests (agentic PRs)}. In parallel, autonomous systems increasingly support the code review process by automatically generating feedback on submitted changes \cite{li2025aiteammates_se3, abdellatif}. These systems, commonly termed as \textit{automated code review agents} \cite{li2022automatingcodereviewactivities} or \textit{reviewer bots} \cite{tufano2024code}, are now frequently involved in reviewing agentic PRs \cite{cihan2025evaluating, abdellatif}, contributing to the broader automation of modern software workflows \cite{bosu, zeng2025benchmarking, GOLZADEH2021110911}. While prior work has examined human perceptions of AI-generated code \cite{schäfer2023empiricalevaluationusinglarge, khatoonabadi}, the interaction between reviewer bots and agentic PRs remains underexplored \cite{watanabe2025useagenticcodingempirical}. In particular, there is limited empirical evidence on how reviewer bot communication characteristics affect workflow outcomes. We argue that understanding the influence of \textit{Bot Feedback Quality} (defined in terms of relevance, clarity, and conciseness), and \textit{Bot Activity Volume} (measured as the number of bot comments per PR), is critical for evaluating review effectiveness, developer experience, and CI/CD pipeline reliability \cite{sun2025does}.

This paper presents a quantitative and correlational analysis of reviewer-bot comments on agentic PRs using review data from the AI\_Dev dataset \cite{li2025aiteammates_se3}. We analyze the types of issues highlighted by bots, the nature and civility of their comments, and assess feedback quality using the categorization and scoring framework proposed by Sghaier et al. \cite{sghaier2025harnessing} (detailed in Section~\ref{sec:sghaier}). We then empirically examine how these communication characteristics relate to PR outcomes. Specifically, we test whether higher \textit{Reviewer Bot Feedback Quality} is associated with reduced PR resolution time and higher acceptance rates, and contrast this effect with that of \textit{Reviewer Bot Activity Volume} \cite{bosu}. Accordingly, we address the following research questions:

\textbf{RQ1:} How can reviewer bot comments on agentic pull requests be characterized in terms of type, nature, civility, and \textit{Bot Feedback Quality} (relevance, clarity, and conciseness)? 

\textbf{RQ2:} To what extent do these variables, specifically the \textit{Bot Feedback Quality} and the \textit{Bot Activity Volume} on a pull request, correlate with PR resolution time and acceptance rate?

\textbf{Replication Kit:} Our replication kit is available at~\cite{zenodo.17866386}. 

\section{Related Work}
Research on automated code review has expanded rapidly, with recent surveys documenting the growing use of large language models (LLMs) in software engineering tasks \cite{widyasari2024explaining}. Prior studies, including those by Schäfer et al. \cite{schäfer2023empiricalevaluationusinglarge} and Tufano et al. \cite{tufano2024code}, evaluate automation for activities such as test generation and code review. While Watanabe et al. \cite{watanabe2025useagenticcodingempirical}, and Wessel et al. \cite{wessel2020, wessel2022} identified this shift; Li et al. \cite{li2025aiteammates_se3} released the AI\_Dev dataset to study it at scale, existing work focuses largely on coarse-grained metrics such as turnaround time and explicitly notes that the dynamics of automated review on agent-generated code remain unexplored. We address this gap by systematically profiling bot comment quality (RQ1) and assessing their correlation with PR resolution and acceptance rates (RQ2). Unlike prior work \cite{khatoonabadi, palvannan2023suggestion, wessel2020effects} that evaluates automated reviewers primarily in terms of their utility to human developers or treats bot feedback as a black box, we quantify specific interaction characteristics, i.e., relevance, clarity, and conciseness, and empirically test whether feedback quality or activity volume drives workflow delays.

\vspace{-0.2cm}
\section{Methodology}

\subsection{Dataset}
We use the AI\_Dev dataset by Li~et~al.~\cite{li2025aiteammates_se3}, which is a large-scale empirical corpus of AI based PRs, review comments, and associated metadata collected from open-source software (OSS) projects on GitHub. We analyze $7,416$ reviewer-bot comments from the AI\_Dev dataset. Each record contains the bot-authored review comment and the associated \texttt{diff\_hunk} (the specific block of modified code). After we pre-process the data and filter out non-bot comments, the final dataset contains $7,416$ reviewer-bot comments generated by $29$ distinct bots across $4,532$ agentic PRs.

\vspace{-0.2cm}
\subsection{Categorization and Scoring Framework}
\label{sec:sghaier}
To identify the interaction patterns of reviewer bots with agentic PRs (RQ1), we adopt the framework by Sghaier et al. \cite{sghaier2025harnessing} and use it to evaluate the characteristics and quality of reviewer-bot comments. Their framework classifies comments along three \textbf{Categorical dimensions:} \textit{Type, Nature, and Civility}, and scores each subcategory according to three \textbf{Scoring Criteria Metrics:} \textit{Relevance, Clarity, and Conciseness}.
Among the categories, \textit{Type} captures focus (e.g., Bugfix, Testing, Documentation), \textit{Nature} captures expression style (e.g., Prescriptive, Clarification), and \textit{Civility} captures tone (Civil or Uncivil) of the comment. Similarly, in the scoring framework, \textit{Relevance} measures pertinence to the code change, \textit{Clarity} measures how clearly the comment conveys its message, and \textit{Conciseness} measures brevity and efficiency, on a scale of 1-10.  A score of $1$ indicates a very irrelevant, unclear, and non-concise comment, and $10$ shows a highly relevant, clear, and concise review comment, respectively.


To annotate reviewer-bot comments, we use GPT-5.1 as an automated annotator following the framework of Sghaier et al. \cite{sghaier2025harnessing}. The model categorizes each comment by \textit{Type}, \textit{Nature}, and \textit{Civility}, and assigns 1–10 ordinal scores for \textit{Relevance}, \textit{Clarity}, and \textit{Conciseness}. The annotation prompt defines all categorization and scoring criteria, includes an illustrative example, and provides full context by incorporating both the review comment text and the associated code changes (diff\_hunk). We first apply this procedure to a random subset of $200$ comments, which are manually validated in parallel. After confirming the reliability of the annotation setup, we apply the automated procedure to the full corpus of $7,132$ reviewer-bot comments. The prompt template is included in our replication kit.

\vspace{-0.2cm}
\subsection{Manual Validation}
To ensure reliability of LLM-generated categories for review comments, two authors independently annotate a random sample of $200$ comments, with a third author resolving the disagreements. For categories, we evaluate inter-rater agreement between human annotators and GPT-5.1 using \textbf{Cohen’s Kappa ($\kappa$)} \cite{cohen}. The $\kappa$ values: Nature ($0.89$), Type ($0.86$), and Civility ($1.0$), indicate almost perfect agreement \cite{Landis1977TheMO}, confirming that GPT-5.1 functions as a reliable high-throughput annotator. For numeric rating dimensions (relevance, conciseness, and clarity on a $1$-$10$ Likert scale), we assess agreement using \textbf{Krippendorff's Alpha ($\alpha$)} with interval-level measurement, which accounts for value-based disagreement \cite{Krippendorff}. To quantify the stability of these estimates, we compute 95\% confidence intervals \cite{Krippendorff_bootstrap}. The results show near-perfect agreement for relevance ($\alpha = 0.94$) and conciseness ($\alpha = 0.95$), and strong agreement for clarity ($\alpha = 0.87$), with narrow confidence intervals confirming the robustness of these estimates. After finding strong inter-rater agreement, We then apply GPT-5.1 to annotate the remaining $7,132$ comments, producing an automated dataset where each comment receives a category, score, and rationale explaining the reasoning behind its classification and scoring.

\vspace{-0.2cm}
\subsection{Correlation Analysis with PR outcomes}
To ensure a fair comparison between accepted and unaccepted PRs for our analysis (RQ2), we establish a consistent observation window to prevent survival bias, where recently submitted PRs might be incorrectly labeled as ``not-merged''. To mitigate this, we define $t_{last}$ as the creation timestamp of the most recently accepted PR and exclude any PRs created after this point, ensuring every PR has sufficient time to be resolved. The final cohort consists of $4,532$ PRs with bot reviews, including $3,054$ accepted and $1,478$ unaccepted cases. We construct the independent variables for RQ2 as follows:

\textbf{1. Bot Feedback Quality Metrics:} We calculate the mean relevance, mean clarity, and mean conciseness of all individual bot comment scores, recorded against a specific PR, as in RQ1.

\textbf{2. Bot Activity Volume Metrics:} We measure the degree of bot engagement volume using the total number of comments on the PR.

We analyze PR outcomes using a two-step non-parametric approach. PR outcomes: \textbf{PR Resolution Time} is defined as the duration between PR creation and acceptance for merged PRs, and \textbf{PR Acceptance Rate} is treated as a binary variable indicating whether a PR was merged (1) or closed without merge (0). First, to assess associations with resolution time, we compute \textbf{Spearman rank correlation coefficients} \cite{spearman} between each independent variable and PR resolution time. It identifies monotonic relationships, specifically testing whether increased \textit{Bot Activity Volume} or improved \textit{Bot Feedback Quality} consistently correlate with faster or slower resolution times. Second, to assess differences in acceptance outcomes, we apply the \textbf{Mann–Whitney U test} \cite{Mann1947OnAT} with 
$\alpha = 0.05$. This test determines if the distributions of \textit{Bot Activity Volume} and \textit{Bot Feedback Quality} scores differ significantly between two independent groups of PRs, i.e., ``accepted PRs'', and ``unaccepted PRs'', thereby identifying bot characteristics associated with successful PR merges. We apply the \textbf{Holm-Bonferroni Correction} \cite{holms} ( a strong variant of Bonferroni) to the p-values of both statistical tests to control the family-wise error rate (FWER) for multiple comparisons.

\section{Results}
This section presents the findings of our research questions, detailing the qualitative coding of the review comments, the quantitative analysis of the distribution of review comments across categories, and the statistical examination of how \textit{Bot Feedback Quality} and \textit{Bot Activity Volume} correlate with PR outcomes.
\subsection{RQ1: Characterizing Reviewer Bot Feedback and Activity}

\subsubsection{Categorization Results} 

Table~\ref{tab:scoring_criteria_averages} presents the categorization of bot comments by \textit{Type, Nature, and Civility}, revealing distinct communication patterns.

\textbf{Type of review comments:} Bugfix ($14.0\%$), Testing ($11.8\%$), and Documentation ($11.6\%$) comments are common, while Refactoring ($5.5\%$) and Logging ($1.7\%$) feedback are rare. The majority of comments ($55.5\%$) fall into the `Other' category, which mainly includes configuration feedback, identified through manual inspection and GPT-5.1-generated rationales.

\textbf{Nature of review comments:} Regarding comment nature, Prescriptive comments constitute a significant share ($32.5\%$), while Clarification ($8.6\%$) and Descriptive ($1.1\%$) comments are minimal, reflecting the non-conversational structure of most bots. Consequently, the `Other' category dominates ($57.8\%$), comprising functional outputs or status notifications that do not fit into standard conversational taxonomies.

\textbf{Civility of review comments:} Bots are overwhelmingly civil: ($99.8\%$) of comments exhibit polite and constructive tone. Uncivil feedback is extremely rare ($0.2\%$) but notable given the potential risks of transferring undesirable language patterns into automated ecosystems. Uncivil comments used the word `useless` for given redundancies in the \texttt{diff\_hunk}.
\footnote{https://github.com/github/gh-gei/pull/1373}

\begin{table}[H]
\centering
\caption{Overall Distribution of Bot Review Categories and Relevance, Clarity, and Conciseness scores per category across the Bot Review Comments in Agentic PRs.}
\resizebox{\columnwidth}{!}{
\label{tab:scoring_criteria_averages}
\begin{tabular}{llcccc}
\toprule
\textbf{Category*} & \textbf{Subcategory*} & \textbf{Category Distribution} &\textbf{Relevance} & \textbf{Clarity} &\textbf{Conciseness} \\
\midrule
\multirow{6}{*}{Type} & Refactoring & 5.5\% & 7.41 & 6.84 & 7.32 \\
& Bugfix & 14.0\% & 7.74 & 6.55 & 6.60 \\
& Testing & 11.8\% & 7.69 & 6.94 & 8.09 \\
& Logging & 1.7\% & 7.21 & 6.86 & 7.56 \\
& Documentation & 11.6\% & 7.39 & 6.67 & 7.38 \\
& Other &55.5\% & 6.59 & 7.07 & 9.25\\
\midrule
\multirow{4}{*}{Nature} & Descriptive &1.1\% & 7.41 & 6.48 & 7.48 \\
& Prescriptive &32.5\% & 7.31 & 7.21 & 7.78 \\
& Clarification &8.6\% & 8.15 & 5.94 & 5.22 \\
& Other & 6.54 & 6.88 & 9.33 \\
\midrule
\multirow{2}{*}{Civility} & Civil &99.8\% & 6.91 & 6.96 & 8.69 \\
& Uncivil &0.2\% & 7.25 & 6.75 & 8.08 \\
\midrule
\multicolumn{3}{c}{\textbf{Average}} & \textbf{6.91} & \textbf{6.96} & \textbf{8.69} \\
\bottomrule
\multicolumn{3}{l}{* From Sghaier et al.\cite{sghaier2025harnessing}} 
\end{tabular}
}
\end{table}

\subsubsection{Scoring Criteria Results} 
Table~\ref{tab:scoring_criteria_averages} summarizes the average scores for each comment category, subcategory (i.e., Type, Nature, Civility). The overall average scores across all comments are $6.91$ for relevance, $6.96$ for clarity, and $8.69$ for conciseness.  The analysis shows that reviewer bots perform best in \textit{Conciseness}, with average scores above $8$ across most categories. \textit{Relevance} and \textit{Clarity} scores are moderate, with \textit{Clarification} comments scoring lowest in conciseness ($5.22$) and clarity ($5.94$), indicating that these comments are less contextually precise and less concise. Among \textit{Type} categories, Other comments achieve the highest conciseness score ($9.25$) but comparatively lower relevance ($6.59$), reflecting functional or automated outputs that are brief but not highly informative. These findings can help developers differentiate between high and moderate value comments, thereby saving their processing time on PRs. Further, bot developers can get insight into how to enhance the effectiveness of automated code reviews by improving the relevance and clarity of bot reviewers.


\vspace{-0.1cm}
\begin{graybox}
\textbf{RQ1 Summary:}
Reviewer-bot comments in agentic PR reviews are predominantly civil, and prescriptive, mainly targeting bug fixes, testing, and documentation. While the comments score high in conciseness ($8.69$), their moderate relevance ($6.91$) and clarity ($6.96$) indicate that developers may be processing many comments that are easy to read but only partially useful, highlighting the need for reviewer bots to prioritize high-value and context-aware feedback.
\end{graybox}

\vspace{-0.3cm}
\subsection{RQ2: What Influences PR Outcomes}
\subsubsection{Impact on PR Acceptance Rate}
\begin{figure}[H]
\centering
\includegraphics[width=1.0\linewidth]{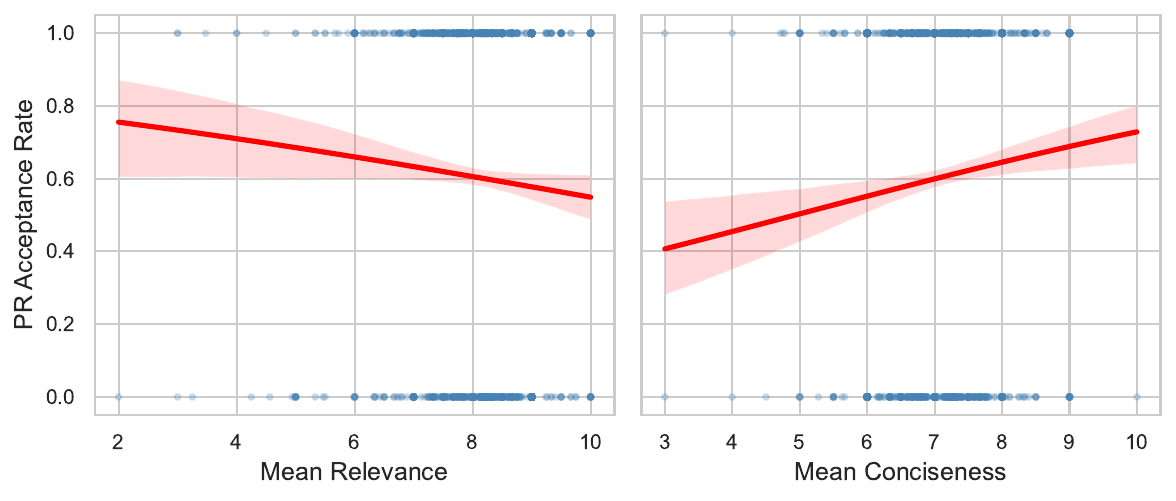}
\caption{\small{Correlations of Mean Relevance and Mean Conciseness of bot comments against PR Acceptance Rate. }}
\label{fig:quality_vs_acceptance_rate}
\end{figure}

The Mann–Whitney U test with $\alpha = 0.05$ confirms that \textit{Bot Feedback Quality} metrics have different distributions when comparing accepted (merged) versus unaccepted (closed without merge) PRs. It indicates that only the mean relevance of bot comments differs significantly between merged and unmerged PRs, after we apply Holm-Bonferroni correction to the p-values ($p_\text{adj} = 0.002$) \cite{holms}, suggesting that higher bot comment relevance is associated with PRs being merged. However, the influence of quality on acceptance is not straightforward, as shown in Figure~\ref{fig:quality_vs_acceptance_rate}. Specifically, \textit{Mean Relevance} shows a significant negative correlation with PR acceptance ($\rho =  -0.09, adj\_p =  0.0006$), suggesting that higher relevance alone does not straightforwardly increase the likelihood of a PR being merged. Conversely, \textit{Mean Conciseness} exhibits a strong positive correlation with PR acceptance ($\rho = 0.06, adj\_p = 0.01$), indicating that concise feedback facilitates a successful merge.

\vspace{-1cm}
\subsubsection{Impact on PR Resolution Time}
Contrary to the hypothesis that high reviewer bot feedback quality would expedite workflows, \textit{Bot Feedback Quality Metrics} show minimal correlation with PR resolution time. In stark contrast, increased \textit{Reviewer Bot Activity Volume}, i.e., bot comments count, shows a significant positive correlation with PR resolution time, as shown in Figure \ref{fig:resolution_time_scatter}. It interprets that PRs receiving more bot comments ($\rho = 0.19, adj\_p = 7.312e-13$) take significantly longer to complete, with p-value values obtained after applying Holm's Bonferroni Correction.
\begin{figure}[H]
\centering
\includegraphics[width=1.0\linewidth]{factors_vs_resolution_time_compact.pdf}
\caption{\small{Correlations of Bot Feedback Quality metrics (Relevance, Clarity, Conciseness) and Bot Activity Volume (Bot Comment Count) against PR resolution time.}}
\label{fig:resolution_time_scatter}
\end{figure}

\subsubsection{The Dilution of Review Quality: Volume vs. Quality}
\begin{figure}[ht]
\centering
\includegraphics[width=1.0\linewidth]{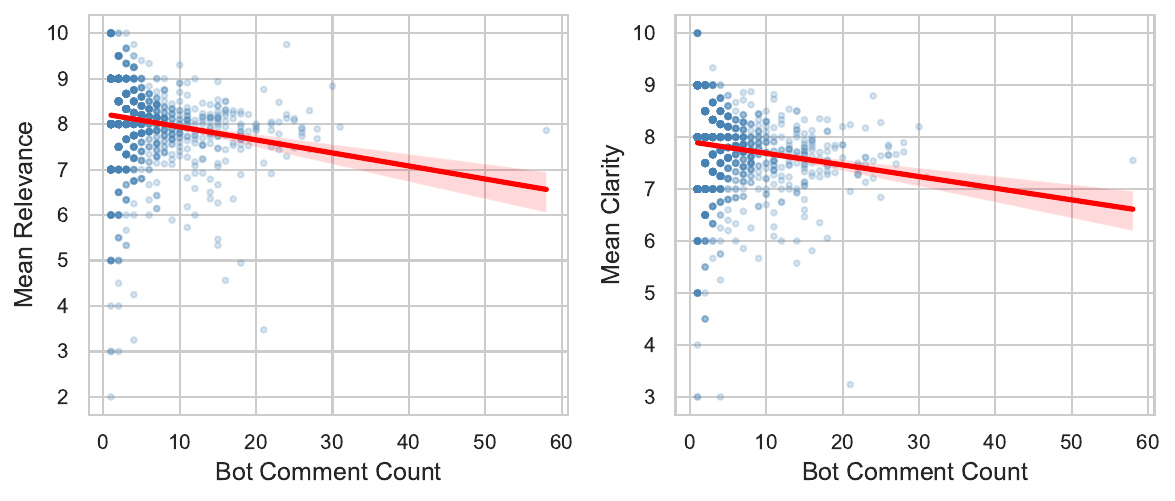}
\caption{\small{Correlation between Bot Comment Count and Mean Relevance of Bot Comments.}}
\label{fig:volume_vs_quality}
\end{figure}

To understand the nature of high-volume reviewer bot activity, we analyze the relationship between \textit{Bot Activity Volume} and \textit{Bot Feedback Quality}. We identify a significant negative correlation between \textit{Bot Comment Count} and \textit{Mean Relevance} ($\rho = -0.2, adj\_p = 8.648e-20$), as well as \textit{Mean Clarity} ($\rho = -0.19, adj\_p = 1.39e-15$). As visualized in Figure \ref{fig:volume_vs_quality}, this inverse relationship indicates a \textit{dilution effect}: as bots generate more comments on a single PR, the average relevance of those comments declines. This suggests that high-volume bots are likely generating `noise', accumulating low-relevance, and less clear messages that degrade the overall review signal.

\vspace{-0.5cm}

\begin{graybox}
\textbf{RQ2 Summary:}
\textit{Bot Feedback Quality} metrics show limited and inconsistent influence on agentic PR outcomes, with relevance and conciseness affecting acceptance in different ways but not accelerating resolution. In contrast, higher \textit{Bot Activity Volume} consistently delays PR completion and reduces average feedback relevance. Overall, workflow efficiency in agentic PRs is driven primarily by comment quantity rather than feedback quality.
\end{graybox}

\vspace{-0.25cm}
\section{Implications}
Our results suggest that reviewer bots on agentic PRs currently function as automated pre-checkers that excel in politeness but may lack the targeted relevance needed for complex workflows. The observed inverse relationship between interaction volume and relevance implies that processing numerous automated comments may become less efficient as their average relevance decreases. This dilution effect, where high-volume bot activity generates low-relevance messages that degrade the overall review signal, suggests that bot design could benefit from evolving toward more strategic assistance. Specifically, implementing interaction thresholds to reduce potentially redundant comments may help mitigate workflow delays. Future research could explore optimal volume thresholds
while developing semantic metrics to better assess the utility of bot feedback relative to the complexity and intent of code changes.

\section{Threats to Validity}
The reviews characterization framework \cite{sghaier2025harnessing} limits construct validity by categorizing $55.5\%$ of bot comments as `Other', which may obscure specific communication subtypes. Our study constrains internal validity by relying on observable comment text and code diffs, thereby preventing analysis of internal bot logic and external developer communications. The exclusive focus on the AI\_Dev dataset and the rapid evolution of agentic tools limit external validity and may affect generalizability to newer ecosystems.

\section{Ethical Implications}
This study does not require formal ethical approval because it does not involve human participants.

\vspace{-0.25cm}
\section{Conclusion}
This study empirically characterizes reviewer-bot feedback in agentic PRs and shows that, while bots consistently produce civil and concise comments, their feedback is only moderately relevant and degrades as activity volume increases. Crucially, our analysis identifies a pattern where \textit{Bot Activity Volume} emerges as the primary driver of longer PR resolution times. We also find that as bots generate more comments, the average relevance and clarity of their feedback decline, indicating a dilution of review signal. Reviewer bots predominantly comment on bug fixing, testing, and documentation, suggesting that their current contribution in agentic workflows aligns with lightweight pre-checking rather than targeted review. These results suggest that reviewer-bot design should emphasize limiting comment volume and prioritizing context-aware feedback generation to mitigate review noise.

\newpage
\bibliographystyle{ACM-Reference-Format}
\bibliography{bibliography}

\end{document}